\documentclass[superscriptaddress,onecolumn]{revtex4}
\usepackage{graphicx,epsfig}
\usepackage{amsmath}
\usepackage {amssymb}
\usepackage{multirow}
\usepackage{float}

\usepackage[dvipsnames]{xcolor}
\RequirePackage[colorlinks,citecolor=blue,urlcolor=blue,linkcolor=blue]{hyperref}

\newcommand{\be}{\begin{eqnarray}}
\newcommand{\ee}{\end{eqnarray}}
\newcommand{\bea}{\begin{eqnarray}}

\newcommand{\eea}{\end{eqnarray}}

\makeindex

\begin{document}



\begin{center}
\large \bf {Quasinormal modes of massive scalar fields in four-dimensional wormholes: anomalous decay rate}
\end{center}


\author{P. A. Gonz\'{a}lez}
\email{pablo.gonzalez@udp.cl} \affiliation{Facultad de
Ingenier\'{i}a y Ciencias, Universidad Diego Portales, Avenida Ej\'{e}rcito
Libertador 441, Casilla 298-V, Santiago, Chile.}

\author{Eleftherios Papantonopoulos}
\email{lpapa@central.ntua.gr}
\affiliation{Physics Division,
National Technical University of Athens, \\15780 Zografou Campus,
Athens, Greece.}

\author{\'Angel Rinc\'on}
\email{angel.rincon@ua.es}
\affiliation{Departamento de Física Aplicada, Universidad de Alicante, Campus de San Vicente del Raspeig, E-03690 Alicante, Spain.}

\author{Yerko V\'asquez}
\email{yvasquez@userena.cl}
\affiliation{Departamento de F\'isica, Facultad de Ciencias, Universidad de La Serena,\\
Avenida Cisternas 1200, La Serena, Chile.}


\begin{abstract}

In this work we consider a Generalized Bronnikov-Ellis wormhole and the tideless Morris-Thorne wormhole and we study the propagation of massive scalar fields. We calculate the quasinormal frequencies using the WKB method and the pseudospectal Chebyshev method and we show the presence of an anomalous decay rate for the quasinormal modes in the Generalized Bronnikov-Ellis wormhole. However, such anomalous behaviour is avoided for the fundamental mode in the  Morris-Thorne wormhole background.

\end{abstract}

\maketitle

\tableofcontents


\clearpage

\section{Introduction}

The recent observations of the relativistic collision of two compact objects produced gravitational waves (GWs) which are probes of the nature of the colliding bodies. Also the recent LIGO detections~\cite{Abbott:2016blz}-\cite{TheLIGOScientific:2017qsa} provided evidence that GW astronomy will help us to understand better  the gravitational interaction and astrophysics in extreme-gravity conditions. However despite   of these developments, the recent observations do not yet probe the detailed structure of spacetime beyond the photon sphere. The future GW observations are expected to detect the ringdown phase, which is governed by a series of damped oscillatory modes at early times, which are known as quasinormal modes (QNMs) \cite{Vishveshwara:1970zz}-\cite{Konoplya:2011qq}, and may potentially contain important information of the structure of compact objects  \cite{Cardoso:2016rao}, specially the physics of the near-horizon region of black holes (BHs) and possible  existence of any unexpected structure.
Future GW observations may give us information on different compact objects than BHs, objects without event horizons. These compact objects are known as
exotic compact objects (ECOs) \cite{Mazur:2001fv}-\cite{Holdom:2016nek}.

The  most well known ECO solutions are the wormholes (WHs) which are  solutions of the Einstein equations that connect different parts of the Universe or two different Universes \cite{MisWheel,Wheel}. Despite that WHs  have different causal structures from BHs, they possess photon spheres and therefore the early stage of the ringdown signal can disguise WHs and  BHs in GW data. Lorentzian wormholes in General Relativity (GR) were discussed in \cite{Bronnikov:1973fh,Ellis:1973yv,Morris,phantom,Visser}, where introducing a static spherically symmetric metric,  conditions for traversable wormholes were found. For these conditions to be satisfied  a matter distribution of exotic or phantom matter, violating the null energy condition (NEC), has to be introduced.  WHs with ordinary matter satisfying the NEC \cite{Visser,Antoniou:2019awm,Mehdizadeh:2015dta} were introduced in modified gravity theories like Brans-Dicke theory \cite{Brans}, $f(R)$ gravity \cite{FR}, Einstein-Gauss-Bonnet theory~\cite{GB}, Einstein-Cartan theory and general scalar-tensor theories \cite{Cartan}. WHs solutions were studied in $f(R)$ theories in \cite{Karakasis:2021tqx}. Also, WHs with a self interacting scalar field were studied in \cite{Anabalon:2012tu}.

Perturbations of BHs and WHs in the Horndeski theory were studied in \cite{Vlachos:2021weq}. In this theory due to the presence of  an effective negative cosmological constant,  AdS asymptotics  are generated due to the presence of a non-minimally coupled scalar to the Einstein tensor. The existence of  AdS asymptotics gives  a different behaviour of the late-time response of the test scalar for BHs and WHs. In the case that the compact object is a BH, the partially reflected waves from the photon sphere (PS)  mirror off the AdS boundary and re-perturb the PS to give rise to a damped beating pattern, while for the case of a WH do not decay with time but have constant and equal amplitude to that of the initial ringdown. This is happening because the reflected waves are related to the absence of dissipation and may be an indication of the existence of normal oscillation modes, as well as potential instabilities.

The above discussion indicates that the knowledge of QNMs and quasinormal frequencies (QNFs) are very important to understand the properties of compact objects and distinguish their nature. The QNMs give an infinite discrete spectrum which consists of complex frequencies, $\omega=\omega_R+i \omega_I$, where the real part $\omega_R$ determines the oscillation timescale of the modes, while the complex part $\omega_I$ determines their exponential decaying timescale (for a review on QNM modes see \cite{Kokkotas:1999bd, Berti:2009kk} and for recent solutions see \cite{Panotopoulos:2020mii,Rincon:2020cos,Rincon:2020pne,Rincon:2020iwy,Panotopoulos:2019gtn,Panotopoulos:2019qjk,Panotopoulos:2020zbj}).

Also, the QNMs and QNFs can give information about the stability of matter fields that evolve perturbatively in the exterior region of a compact object without backreacting on the metric. For the case of Schwarzschild and Kerr black hole background it was found that the longest-lived modes are always the ones with lower angular number $\ell$. This is be understood from the fact that  the more energetic modes with high angular number $\ell$ would have faster decaying rates.

If the probe scalar field is massive, then a mass scale is introduced and  a different behaviour  was found
\cite{Konoplya:2004wg, Konoplya:2006br,Dolan:2007mj, Tattersall:2018nve},  at least for the overtone $n = 0$. If the mass of the scalar field is  light, then the longest-lived QNMs are those with a high angular number $\ell$, whereas if the mass of the scalar field is large  the longest-lived modes are those with a low angular number $\ell$. This behaviour is expected since if the probe scalar field is massive its fluctuations can maintain the QNMs to live longer even if the angular number $\ell$ is large. This behaviour of the QNMs introduces an anomaly of the decaying modes which depends on whether the mass of the scalar field exceeds a critical value or not. Introducing another scale in the theory through the presence of a cosmological constant an anomalous behaviour of QNMs was found in \cite{Aragon:2020tvq}  as the result of the  interplay of the mass of the scalar field and the value of the cosmological constant. Anomalous QNMs decay modes were  also found if the background metric is the Reissner-Nordstr\"om and the probe scalar field is massless \cite{Fontana:2020syy} of massive \cite{Gonzalez:2022upu} depending on critical values of the charge of the black hole, the charge of the scalar field and its mass. The decay modes of QNMs in various setups were studied in \cite{Aragon:2020teq}-\cite{Aragon:2021ogo}.

Recently there are studies of QNMs and QNFs of other compact objects like WHs. In the case of BHs to calculate the QNFs of the scalar field equation we assume a purely outgoing wave at infinity and a purely incoming wave at the event horizon. However, a transformation to tortoise coordinate changes the boundary conditions of the scalar field, there are no incoming waves from both plus- and minus- infinity. The traversable WHs that  connect two infinities in terms of the tortoise coordinate have the same boundary conditions and therefore the QNM spectrum of WHs should have some similarities and differences with the BHs spectrum. The evolution of perturbations of WH spacetime was investigated in \cite{Konoplya:2005et}  and it was found that the QNMs are behaving the same way as a BH, includes the quasinormal ringing and power-law asymptotically late time tails. However, in \cite{Konoplya:2016hmd} it was found that the symmetric WHs do not have the same ringing behaviour of the BHs  at a few various dominant multipoles. In \cite{Churilova:2019qph} perturbations of a massive probe scalar field were considered and  it was shown that WHs with a constant redshift function do not allow for  longest-lived modes, while wormholes with non-vanishing radial tidal force do allow for quasi-resonances. In the Morris-Thorne wormhole spacetime calculating the high frequency QNMs  modes the shape function of a spherically symmetric traversable Lorenzian wormhole near its throat was constructed in \cite{Konoplya:2018ala}. Scalar and axial QNMs of massive static phantom WHs were calculated in \cite{Blazquez-Salcedo:2018ipc}.

As we have already discussed, in the case of BHs, if the probe scalar field is massive then an anomalous behaviour of QNMs was observed. In this work we will investigate if the same behaviour is observed in the case that the background metric defines a WH geometry. We will work with two WHs solutions, the Generalized Bronnikov-Ellis wormhole \cite{Ellis:1973yv} and the Morris-Thorne wormhole \cite{Morris:1988tu}. For the case of the Generalized Bronnikov-Ellis wormhole we will show that there is a critical mass of the scalar field beyond which the anomalous decay rate of the QNMs is present.
In the case of the  Morris-Thorne wormhole the anomalous decay rate of the QNMS is not present at least for the fundamental modes. We will also show that
the propagation of massive scalar fields is stable in these wormhole geometries.

Our work in the present article is organized as follows: In Section \ref{sec2} we  review   the equations of traversable wormholes considered. In Section \ref{sec3} we study the massive scalar field perturbations. In Section \ref{sec4} we obtain the QNFs by using the WKB method and the pseudospectral Chebyshev method and we analyze the anomalous decay rate of the QNFs. Finally, in Section \ref{sec5} we summarize our work with some concluding remarks.

\section{Review of  traversable wormholes}

\label{sec2}

Considering a  static spherically symmetric metric the necessary conditions to generate traversable Lorentzian wormholes as  exact solutions in GR were first found  by Morris and Thorne \cite{Morris}. To find a solution which is consistent with a condition on the wormhole throat necessitates the introduction of   exotic matter, which   leads to the violation of the NEC.

We consider the following action
\begin{equation} S = \int d^4x \sqrt{-g} \left( \frac{R}{2} + \frac{1}{2}\partial^{\mu}\phi\partial_{\mu}\phi -V(\phi) \right) ~,\label{action} \end{equation}
which consists of  the Ricci Scalar $R$, a scalar field with negative kinetic energy and a self-interacting potential.

The field equations that arise by variation of this action  are
\begin{eqnarray}
R_{\mu\nu} - \frac{1}{2}g_{\mu\nu}R  &=& T_{\mu\nu}^{\phi}~, \label{EE}\\
\Box \phi + V_{\phi}(\phi) &=&0~, \label{KG}
\end{eqnarray}
where $\Box = \nabla^{\mu}\nabla_{\mu}$ is the D'Alambert operator with respect to the metric,  and the energy-momentum tensor is given by
\begin{equation} T_{\mu\nu}^{\phi} = - \partial_{\mu}\phi \partial_{\nu}\phi + \frac{1}{2}g_{\mu\nu}\partial^{\alpha}\phi \partial_{\alpha}\phi - g_{\mu\nu}V(\phi)~. \end{equation}
We consider the following metric ansatz firstly used by  Morris and Thorne \cite{Morris} in spherical coordinates
\begin{equation} ds^2 = - e^{2\Phi(r)}dt^2 + \left(1-\frac{b(r)}{r}\right)^{-1} dr^2 + r^2 d\Omega\label{MorrisThorne-metric} ~,\end{equation}
where $\Phi(r)$ is the redshift function, $b(r)$ is the shape function and $d\Omega = d\theta^2 + \sin(\theta)^2d\varphi^2$. In order to obtain a wormhole geometry, these functions have to satisfy the following conditions \cite{Morris,Visser}, namely:
\begin{enumerate}
    \item $\frac{b(r)}{r}\leq 1$ for every $[r_{th},+\infty)$, where $r_{th}$ is the radius of the throat. This condition ensures that the proper radial distance defined by $l(r)=\pm\int_{r_{th}}^{r}\left(1-\frac{b(r)}{r}\right)^{-1} dr$ is finite everywhere in spacetime.
   Note that in the coordinates $(t,l,\theta,\phi)$ the line element \eqref{MorrisThorne-metric} can be written as \begin{align}
    ds^2=-e^{2\Phi(l)}dt^2+dl^2+r^2(l)(d\theta^2+\sin^2\theta d\varphi)\,.\label{metric-proper-chart}
    \end{align}
   In this case the throat radius would be given by $r_{th}=\min\{r(l)\}$.
    \item $\frac{b(r_{th})}{r_{th}}=1$ at the throat. This relation comes from requiring the throat to be a stationary point of $r(l)$. Equivalently, one may arrive at this equation by demanding the embedded surface of the wormhole to be vertical at the throat.
    \item $b'(r)<\frac{b(r)}{r}$ which reduces to $b'(r_{th})\leq 1$ at the throat. This is known as the flare-out condition since it guarantees $r_{th}$ to be a minimum and not any other stationary point.
    \end{enumerate}

The ensure the absence of horizons and singularities requires  $\Phi(r)\neq 0$  which means that $\Phi(r)$ is finite throughout the spacetime. To be more specific the Ellis \cite{Ellis:1973yv} is a wormhole solution of an action that consists of a pure Einstein-Hilbert term and a scalar field with negative kinetic energy
\begin{equation}
	S = \int d^4x \sqrt{-g} \left(\frac{R}{2} + \frac{1}{2}\partial^{\mu}\phi\partial_{\mu}\phi\right)~.
\end{equation}
Assuming the metric ansatz (\ref{MorrisThorne-metric}), for $f(r)=R(r),  V(r)=0$  the above equations give the following solution
\begin{eqnarray}
b(r) &=&A^2/r~, \label{bellis}\\
\phi(r) &=& \sqrt{2} \tan ^{-1}\left(\frac{\sqrt{r^2-A^2}}{A}\right)+\phi_{\infty}~, \label{ellisscalar}\\
f(r) &=& R(r) =-\frac{2 A^2}{r^4}~, \label{rellis}
\end{eqnarray}
where $A, \phi_{\infty}$ are two constants of integration. The resulting spacetime is asymptotically flat as it can be seen from both $b(r)$ and $R(r)$. The wormhole throat is the solution of the equation
\begin{equation} g_{rr}^{-1} = 0 \rightarrow r_{th} = \pm A~. \end{equation}
The solution also satisfies the flaring-out condition and $b'(r_{th})=-1$.
\newline

The scalar field takes a constant value at infinity $\phi(r\to \infty) = \frac{\pi }{\sqrt{2}} +\phi_{\infty} $, so to make the scalar field vanish at large distances we will set $\phi_{\infty} = -\frac{\pi }{\sqrt{2}}$.  The scalar field takes the asymptotic value at infinity at the position of the throat $\phi(r=r_{th})=\phi_{\infty}$.  As can be seen in the solution (\ref{bellis}), (\ref{ellisscalar}) and (\ref{rellis}) the integration constant $A$  of the phantom scalar field plays a very important role in the formation of the wormhole geometry.  It has  units $[L]^2$, appears in the scalar curvature and at the position of the throat takes the value $R_{r = r_{\text{th}}}(A) = -2/A^2$. Also it effects the size of the throat, a larger charge $A$ gives a larger wormhole throat. Therefore, the presence of the phantom scalar field is very important for the generation of the wormhole geometry and to define the scalar curvature and specifying  the throat of the wormhole geometry.

In addition, the function $b(r)$ encodes the shape of the wormhole. Notice that  at certain minimum value of $r$, the throat of a wormhole is defined, namely, when  $r_{min} = b_0$. Thus,  the radial coordinate increases ranging from $r_{min}$ until spatial infinity $r = \infty$. On one hand, keep in mind that  $\Phi(r)$ must be finite everywhere in light of the requirement of absence of singularities. In addition, $\Phi(r) \rightarrow 0$ as $r \rightarrow \infty $ (or $ l \rightarrow \pm \infty$) based on the requirement of asymptotic flatness.
On the other hand, the shape function $b(r)$ should satisfy that $1- b(r)/r > 0$ and
$b(r)/r \rightarrow 0$ as $r \rightarrow \infty $ (or equivalently $ l \rightarrow \pm \infty$).
In the throat $r = b(r)$ and thus $1- b(r)/r$ vanishes.
Traversable wormholes does not have a singularity in the throat. The later means that travellers can pass through the wormhole during the finite time.

In particular, we shall consider the following metric functions
\begin{equation} \label{wormhole}
    e^{2 \Phi(r)} = 1 -\frac{2M}{r}\,, \,\,\,\, b(r)= \frac{b_0^2}{r} \,,
\end{equation}
which goes over into the Bronnikov-Ellis wormhole in the limit $M \rightarrow 0$ \cite{Bronnikov:1973fh,Ellis:1973yv},
and we also consider the  Morris-Thorne wormhole, characterized by $b(r) = \sqrt{b_0 r }$. Note that the parameter $b_0$ specifies the charge of the phantom scalar field.
\newline

The most general energy-momentum tensor compatible with the wormholes geometries is given by \cite{Bronnikov:2018vbs}
\begin{equation}
    T^{\mu}_{\,\,\,\,\, \nu} = diag(- \rho, p_r, p_t, p_t) \,,
\end{equation}
where $\rho$ is the energy density and $p_r$, $p_t$ are radial and tangential pressures respectively.

The Einstein's field equations take the form
\begin{eqnarray}
\nonumber \rho &=& \frac{b'(r)}{r^2} \,,  \\
\nonumber p_r &=& \frac{-b(r)+2 r (r-b(r)) \Phi '(r)}{r^3} \,,  \\
\nonumber p_t &=& \frac{(b(r)- r b'(r) +2 r (r-b(r)) \Phi'(r))(1+ r \Phi'(r)) + 2 (r - b(r))r^2 \Phi''(r)}{2 r^3} \,.
\end{eqnarray}
The important energy conditions are the weak energy condition (WEC), the null energy condition (NEC), the strong energy condition (SEC) and the dominant energy condition (DEC). In terms of the principal pressures the energy conditions are given by, see for instance \cite{Samanta:2019tjb}
\begin{eqnarray}
  \nonumber  && \text{WEC}: \rho \geq 0, \rho + p_r \geq 0, \rho + p_t \geq 0 \\
  \nonumber  && \text{NEC}: \rho + p_r \geq 0, \rho+ p_t \geq 0 \\
   \nonumber && \text{SEC}: \rho+ p_r \geq 0, \rho+ p_t \geq 0, \rho+ p_r+ 2 p_t \geq 0 \\
  \nonumber  && \text{DEC}: \rho \geq 0, \rho - |p_r| \geq 0, \rho - |p_t| \geq 0
\end{eqnarray}
For the generalized Bronnikov-Ellis wormhole we obtain
\begin{eqnarray}
 \nonumber   \rho &=& -\frac{b_0^2}{r^4}\,,  \\
 \nonumber   p_r &=& \frac{-b_0^2 + 2M r}{r^3(r-2M)} \,, \\
 \nonumber   p_t &=& \frac{(b_0^2 (M-r)+Mr^2)}{r^4 (r- 2M)^2} \,, \\
  \nonumber  \rho+ p_r &=& \frac{2(b_0^2(M-r)+ M r^2)}{r^4 (r- 2M)} \,, \\
  \nonumber  \rho+ p_t &=& \frac{M (b_0^2 (2 r -3M)+ (M-r)r^2)}{r^4 (r-2M)^2}\,.
\end{eqnarray}
Therefore $\rho < 0$ and the WEC is violated for any $r > r_{th} = b_0$. Also, $\rho + p_r$ is negative in the range $b_0 \leq r < \frac{b_0^2+\sqrt{b_0^4- 4b_0^2 M^2}}{2 M}$. On the other hand, $p_t$ is negative for $r > \frac{b_0^2+\sqrt{b_0^4- 4b_0^2 M^2}}{2 M}$, therefore this implies that $\rho+ p_t$ at least is negative in that range. Therefore all the energy conditions are violated for $r> r_{th}$.

For the Morris-Thorne wormhole $\rho = \frac{\sqrt{b_0 r}}{2 r^3} >0$; however, $\rho+ p_r = -\frac{\sqrt{b_0 r}}{2 r^3} <0$ and we can verify that all the energy conditions are also violated. 

\section{Massive scalar fields perturbations}
\label{sec3}

In order to study the propagation of a massive scalar field in a wormhole geometry we consider  the  Klein-Gordon equation given by
\begin{equation}
\frac{1}{\sqrt{-g}} \partial_\mu (\sqrt{-g} g^{\mu \nu} \partial_\nu) \Psi = m^2 \Psi\,.
\end{equation}
So, to decouple and subsequently resolve the Klein-Gordon equation, we take advantage of the method of separation of variables making the following Ans{\"a}tz:
\begin{equation}\label{separable}
\Psi(t,r,\theta,\phi) = e^{-i \omega t} \: \frac{y(r)}{r} \: \tilde{Y}_l (\Omega)\,,
\end{equation}
where $\omega$ is the unknown frequency (which will be determined), while $\tilde{Y}_l(\Omega)$ are the spherical harmonics, and they depend on the angular coordinates only \cite{book}.

After the implementation of the above mentioned  Ans{\"a}tz it is easy to obtain, for the radial part, a Schr{\"o}dinger-like equation, namely
\begin{equation}
\frac{\mathrm{d}^2 y}{\mathrm{d}r^{\ast}{}^2} + [ \omega^2 - V(r^{\ast}) ] y = 0\,,
\label{sequation}
\end{equation}
where $r^{\ast}$ is the well-known tortoise coordinate, i.e.,
\begin{equation}
r^{\ast}  \equiv  \int \frac{\mathrm{d}r}{e^{\Phi} \sqrt{1 -\frac{b(r)}{r}}}\,.
\end{equation}
Finally, the effective potential (for scalar perturbations) is then given by \cite{ref}
\begin{equation}
V(r) = e^{2\Phi} \: \left(m^2 + \frac{\ell (\ell+1)}{r^2}
-
\frac{b' r - b}{2r^3}
+
\frac{1}{r}
 \Bigg(1 -\frac{b'}{r}\Bigg) \Phi'
 \right)\,,
\label{pot}
\end{equation}
where, the prime denotes differentiation with respect to the radial coordinate, and $\ell  \geq 0$ is the angular degree. Be aware and notice that we can identify $L^2 \equiv \ell(\ell + 1)$ as  the square of the angular momentum. Thus, we have reduced the problem to the familiar one dimensional Schr{\"o}dinger equation with energy $\omega^2$ and effective potential $V(r)$.
%
%
%

%

The wormhole connects two different regions of the space. Such regions are located at "$\pm \infty$", and only via the throat is where such spaces connect. 
%
%
%
%
In order to obtain the QNFs, it is necessary to consider appropriated boundary conditions. The wormhole geometry is asymptotically flat and the effective potential (with respect to the tortoise coordinate) is a barrier which
take constant values at both infinities, therefore we will consider as boundary conditions the requirement of purely outgoing waves at both infinities: no waves are coming from both asymptotically flat regions \cite{Konoplya:2005et,Bronnikov:2021liv}
\begin{align}
\nonumber  &\Psi \sim e^{ + i \sqrt{\omega^2-m^2} r^{\ast}}\,, \quad r^{\ast} \rightarrow + \infty\,.
\\
&\Psi \sim e^{ - i \sqrt{\omega^2-m^2} r^{\ast}}\,, \quad r^{\ast} \rightarrow - \infty\,.
\end{align}
\section{Quasinormal frequencies}

\label{sec4}

\subsection{Numerical methods}
\subsubsection{WKB method}

The semi-analytical Wentzel-Kramers-Brillouin (WKB) method can be used for effective potentials which have the form of potential barriers that approach to constant values at the horizon and at spatial infinity \cite{Konoplya:2011qq} and has been used to determine the QNFs for asymptotically flat and asymptotically de Sitter black holes, also it has been applied to determine the QNFs of asymptotically flat wormhole geometries. The QNMs are determined by the behaviour of the effective potential near its maximum value $r^*_0$. The Taylor series expansion of the potential around its maximum is given by
\begin{equation}
V(r^*)= V(r^*_0)+ \sum_{i=2}^{\infty} \frac{V^{(i)}}{i!} (r^*-r^*_0)^{i} \,,
\end{equation}
where
\begin{equation}
V^{(i)}= \frac{d^{i}}{d r^{*i}}V(r^*)|_{r^*=r^*_0}
\end{equation}
corresponds to the $i$-th derivative of the effective potential with respect to the tortoise coordinate $r^*$ evaluated at the maximum of the potential. The QNFs in the WKB approximation carried to third order beyond the eikonal approximation are given by \cite{Hatsuda:2019eoj}
\begin{eqnarray} \label{frecuencia}
\omega^2 &=& V(r^*_0)  -2 i U \,,
\end{eqnarray}
where
\begin{eqnarray}
\notag U &=&  N\sqrt{-V^{(2)}/2}+\frac{i}{64} \left( -\frac{1}{9}\frac{V^{(3)2}}{V^{(2)2}} (7+60N^2)+\frac{V^{(4)}}{V^{(2)}}(1+4 N^2) \right) +\frac{N}{2^{3/2} 288} \Bigg( \frac{5}{24} \frac{V^{(3)4}}{(-V^{(2)})^{9/2}} (77+188N^2) + \\
\notag && \frac{3}{4} \frac{V^{(3)2} V^{(4)}}{(-V^{(2)})^{7/2}}(51+100N^2)  +\frac{1}{8} \frac{V^{(4)2}}{(-V^{(2)})^{5/2}}(67+68 N^2)+\frac{V^{(3)}V^{(5)}}{(-V^{(2)})^{5/2}}(19+28N^2)+\frac{V^{(6)}}{(-V^{(2)})^{3/2}} (5+4N^2)  \Bigg)   \,,
\end{eqnarray}
and $N=n+1/2$, where $n=0,1,2,\dots$ is the overtone number. Now, by considering $L^2 = \ell (\ell+1)$ large, the QNFs are approximately given by
\begin{equation} \label{aproximado}
\omega = \omega_1 L + \omega_0 +\omega_{-1}L^{-1} + \omega_{-2}L^{-2} + \mathcal{O}(L^{-3}) \,.
\end{equation}
In the next section we shall employ the third order WKB formula (\ref{frecuencia}) to determine approximate analytical values of the QNFs for large values of the angular momentum (\ref{aproximado}) and the critical scalar field mass for the two wormhole geometries mentioned. Additionally,
we have used
the Wolfram Mathematica \cite{wolfram} code utilizing the WKB method at order six \cite{Konoplya:2019hlu} to evaluate the QNFs numerically.

%

\subsubsection{Pseudospectral Chebyshev method}

In this section we use the pseudospectral Chebyshev method to calculate the QNFs for a massless scalar field in the Generalized Bronnikov-Ellis wormhole.
In this case, Eq. (\ref{sequation}) can be written as
\begin{eqnarray}  \label{nueva}
\nonumber     (2 M-r) \left(b_0^2-r^2\right) y''(r)+ && \frac{\left(-3 M b_0^2+M r^2+b_0^2 r\right)}{r}y'(r)+  \\
    && \frac{\left(M \left(\left(2 \ell^2+2 \ell-1\right) r^2+3 b_0^2\right)-r \left(r^2 \left(\ell^2+\ell-r^2 \omega ^2\right)+b_0^2\right)\right)}{r^2} y(r) =0\,.
\end{eqnarray}
The effective potential $V(r^{\ast}) \rightarrow 0$ when $r^{\ast} \rightarrow \infty$ and the boundary condition satisfied by the QNMs at infinity is given by
\begin{equation}
\nonumber   y \sim e^{i \omega r^{\ast}}   \,\,\,  \text{as} \,\,\,   r^{\ast} \rightarrow \infty\,.
\end{equation}
So, there are only outgoing waves at infinity, which can be transformed to
\begin{equation} \label{bc}
y \sim  e^{i \omega r} r^{i \omega M} \,\,\,    \text{as} \,\,\, r \rightarrow \infty\,.
\end{equation}
Also, since the potential is symmetric about the throat of the wormhole, which is located at $r_0^{\ast}=0$, the solutions will be symmetric or anti-symmetric. Therefore, we impose the following boundary conditions at the throat, $\frac{dy}{d r^{\ast}}\left|_{r_0^{\ast}}=0\right.$ for the symmetric solutions and $y(r_0^{\ast}) = 0$ for the anti-symmetric solutions. These boundary conditions yields the even and odd overtones, respectively. Similar boundary conditions have been applied in \cite{Aneesh:2018hlp, DuttaRoy:2019hij} for other wormhole geometries, where the QNMs were obtained by direct integration of the wave equation.
\newline

For the symmetric solutions it is convenient to define $y = e^{i \omega r} r^{i \omega M} \chi(r)$, which satisfies the boundaries conditions when $\chi(r)$ is regular in $b_0 \leq r < \infty$. Then, performing the following change of coordinate $z =1-\frac{b_0}{r}$, Eq. (\ref{nueva}) is transformed to
\begin{eqnarray}
 \nonumber  && (z-2) z (z-1)^2\left(b_0+2 M (z-1)\right) \chi ''(z)+ \Big((z-1) \left(3 b_0 (z-2) z+b_0+M (z-1) (7 (z-2) z+2)\right)+   2 i \omega  (z-2) \\
 \nonumber && z \left(b_0+2 M (z-1)\right) \left(b_0- M z+M\right)\Big)\chi '(z)+  \Big(b_0 (-b_0^2 \omega ^2+i b_0 \omega  (z-1) +\ell^2+\ell+(z-1)^2 ) +M (z-1) (i b_0 \omega  \\
 \nonumber   &&  (z-1)  +   2 \ell (\ell+1)+3 z^2-6 z+2)+M^2 \omega  (3 b_0 \omega  (z-2) z-  i (z-1) (5 (z-2) z+2)) -2 M^3 \omega ^2 (z-2) \\
  &&  (z-1) z\Big)\chi (z) =0 \,,
\end{eqnarray}
where the coordinate $z$ lies in the range $[0,1]$, $z=0$ corresponds to the throat and $z=1$ corresponds to spatial infinity.
\newline

For the anti-symmetric solutions it is convenient to define $y = (r-b_0)^{1/2} e^{i \omega r} r^{i \omega M-1/2} \chi(r)$, which satisfies the boundaries conditions when $\chi(r)$ is regular in $b_0 \leq r < \infty$. Then, performing the following change of coordinate $z =1-\frac{b_0}{r}$, Eq. (\ref{nueva}) is transformed to
\begin{eqnarray}
\nonumber && 4 b_0^2 (z-2) z (z-1)^2 \left(b_0+2 M (z-1)\right) \chi ''(z)+4 b_0^2  \Big((z-1) \left(b_0 (z (4 z-9)+3)+M (z-1) (z (9 z-20)+6)\right)+  \\
\nonumber && 2 i \omega  (z-2) z \left(b_0+2 M (z-1)\right) \left(b_0+M (-z)+M\right)\Big)\chi '(z)+b_0^2  \Big(2 M (z-1) (2 i b_0 \omega  (2 z-3)+4 \ell (\ell+1) +12 z^2 - \\
\nonumber  && 29 z+15)+b_0 (-4 b_0^2 \omega ^2+4 i b_0 \omega  (2 z-3)+4 \ell (\ell+1)+9 z^2-22 z+13)+4 M^2 \omega  (3 b_0 \omega  (z-2) z -i (z-1) (z   \\
&&
(7 z-16)+6))-8 M^3 \omega ^2 (z-2)(z-1) z\Big)\chi (z) =0\,.
\end{eqnarray}

The pseudospectral method is implemented by expanding $\chi (z)$ in a complete basis of functions $\{\varphi_i(z)\} $: $\chi (z)=\sum_{i=0}^{\infty} c_i \varphi_i(z)$, where $c_i$ are the coefficients of the expansion, and we choose the Chebyshev polynomials $T_j(x)=\cos (j \cos^{-1}x)$ as the basis that are well defined in the interval $x \in [-1,1]$, and $j$ corresponds to the grade of the polynomial. The sum is truncated until some value $N$, therefore the function $\chi (z)$ is approximated by
\begin{equation}
\chi (z) \approx \sum_{i=0}^N c_i T_i (x)\,.
\end{equation}
Thus, the solution is assumed to be a finite linear combination of the Chebyshev polynomials. Since $z \in [0,1]$, the relation between the coordinates $x$ and $z$ is $x=2z-1$. Then, the interval $[0,1]$ is discretized at the Chebyshev collocation points $z_j$ by using the so-called Gauss-Lobatto grid, where
\begin{equation}
    z_j=\frac{1}{2}[1-\cos(\frac{j \pi}{N})]\,, \,\,\,\, j=0,1,...,N \,.
\end{equation}
The corresponding differential equation is then evaluated at each collocation point. So, a system of $N+1$ algebraic equations is obtained, which corresponds to a generalized eigenvalue problem which is solved numerically to obtain the QNMs spectrum, by employing the built-in Eigensystem [ ] procedure in Wolfram’s Mathematica \cite{wolfram}.

In Tables \ref{TableI} and \ref{TableII} we show the fundamental QNFs and the first overtone calculated using the WKB approximation and the pseudospectral Chebyshev method, we see a good agreement, specially for high $\ell$. We observe that when $M$ increases the WKB method is less accurate for low values of $\ell$.


\begin{table}[H]
\centering
\caption{QNFs $\omega b_0$ for massless scalar field perturbations for the Bronnikov-Ellis wormhole with $n=0,1$, for several values of the angular momentum $\ell$ of the scalar field using the six order WKB approximation and the pseudospectral Chebyshev method. }
\begin{tabular}{|c|c|c|}\hline
\multicolumn{3}{|c|}{$n=0$} \\ \hline
$\ell$ &  WKB & pseudospectral  \\
\hline
  1  & 1.58240688-0.52094760 i &  1.57270839 - 0.52970088 i \\
  3  &  3.53418275-0.50723853 i  & 3.53428741-0.50692165 i \\
  5  & 5.52229633-0.50299528 i & 5.52230920-0.50295823 i  \\
  10   & 10.51183943-0.50083950 i & 10.51183964-0.50083853 i \\
    15  & 15.50804387-0.50038776 i &  15.50804389 - 0.50038767 i   \\

   20  & 20.50608858-0.50022226 i & 20.50608858 - 0.50022224 i \\
  30  & 30.50409562-0.50010061 i &  30.50409562-0.50010061 i \\
\hline
\multicolumn{3}{|c|}{$n=1$} \\ \hline
$\ell$ &  WKB & pseudospectral  \\
\hline
  1  & 1.26019632-1.65600690 i & 1.25582759 -
 1.70245148 i  \\
  3  &  3.39009551-1.53845539 i  & 3.39009976-1.53721055 i  \\
  5  & 5.43086691 - 1.51543326 i & 5.43090081-1.51524731 i  \\
  10   &  10.46413272-1.50423343 i  & 10.46413347 - 1.50422829 i \\
    15  & 15.47575749-1.50194642 i &  15.47575755 - 1.50194589 i \\
   20  & 20.48168591-1.50111377 i & 20.48168592 - 1.50111367 i \\
  30  & 30.48769836-1.50050355 i & 30.48769836 - 1.50050354 i  \\
\hline

\end{tabular}
\label{TableI}
\end{table}



\begin{table}[H]
\centering
\caption{QNFs $\omega b_0$ for massless scalar field perturbations for the Generalized Bronnikov-Ellis wormhole with $M/b_0 = 0.2$,  $n=0,1$, for several values of the angular momentum $\ell$ of the scalar field using the six order WKB approximation and the pseudospectral Chebyshev method. }

\begin{tabular}{|c|c|c|}
\hline
\multicolumn{3}{|c|}{$n=0$} \\ \hline
$\ell$ &  WKB & pseudospectral  \\
\hline
  1  & 2.07409974-0.01995352 i &  1.26184288-0.35838100 i \\
   3  &  2.78480392-0.31403523 i & 2.76272243-0.32940500 i  \\
  5  & 4.29701216-0.32089962 i & 4.29516650-0.32260412 i  \\
  10   & 8.15242428-0.31819085 i & 8.15238600-0.31824940 i \\
    15  & 12.01936574-0.31718861 i & 12.01936256-0.31719541 i \\

   20  & 15.88919723-0.31678931 i & 15.88919672-0.31679069 i \\
  30  & 23.63192203-0.31648546 i & 23.63192200-0.31648560 i  \\
\hline
\multicolumn{3}{|c|}{$n=1$} \\ \hline
$\ell$ &  WKB & pseudospectral  \\
\hline
 1  & 5.60190637+0.44619248 i & 1.12961440-1.14814627 i  \\
  3  &  2.92789699-0.81127205 i  &  2.71575841-1.00573220 i \\
  5  & 4.29045541-0.95572400 i & 4.27080139-0.97685590 i  \\
  10   &  8.14306669-0.95728640 i  & 8.14260660-0.95800302 i \\
    15  & 12.01345291-0.95314405 i &  12.01341265-0.95322867 i \\
   20  & 15.88492607-0.95133409 i & 15.88491952-0.95135153 i \\
  30  & 23.6291682-0.94991320 i & 23.62916774-0.94991495 i \\ \hline
\end{tabular}
\label{TableII}
\end{table}

\begin{figure}[h]
\begin{center}
\includegraphics[width=0.5\textwidth]{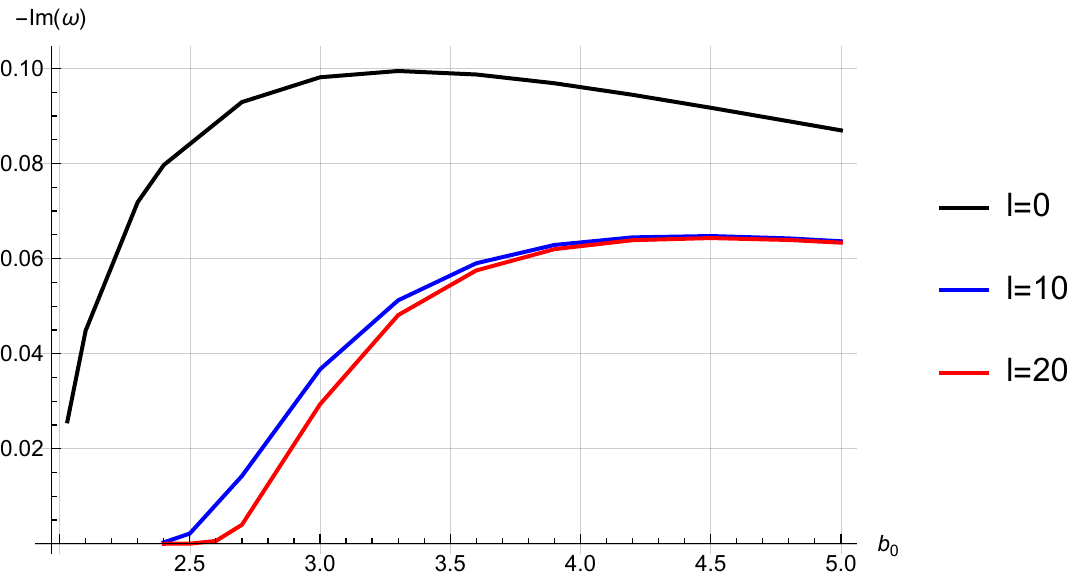}\\
\end{center}
\caption{The behaviour of $- Im(\omega)$ for massless scalar field as a function of $b_0$ for the Generalized Bronnikov-Ellis wormhole with $M=1$, $n=0$, and $\ell = 0, 10, 20$ using the pseudospectral Chebyshev method.}
\label{omega}
\end{figure}

In Table \ref{Table} we show the fundamental QNFs for several values of $\ell$ and $M/b_0$ using the pseudospectral Chebyshev method.  We observe that the imaginary part of the QNFs decreases as $\ell$ increases; however, for $M/b_0=0.4$ the decreasing is much faster and the imaginary part tends to zero as $\ell$ increases.  Furthermore, in Fig.~\ref{omega} we plot the behaviour of the imaginary part of the QNFs for a massless scalar field as a function of $b_0$ for $M=1$ and $\ell=0, 10, 20$. Note that the imaginary part tends to zero when $b_0 \rightarrow 2 M$, and there is a range where $-Im(\omega)$ increases when $b_0$ increases until it reaches a maximum value and then begin to decreasing when $b_0$ increases. Also, for a fixed value of $b_0$ we observe that $-Im(\omega)$ decreases when $\ell$ increases.

\begin{table}[H]
\centering
\caption{Fundamental QNFs $\omega b_0$ for massless scalar field perturbations for the Generalized Bronnikov-Ellis wormhole for several values of the angular momentum and $M/b_0$ using the pseudospectral Chebyshev method.}

\begin{tabular}{|c|c|c|c|c|c|}\hline
 & $M/b_0 =0.1$ &

 $M/b_0 =0.2$ &  $M/b_0 =0.3$  & $M/b_0 =0.4$ \\ \hline
 $\ell=0$ & 0.64217472-0.52979187 i  &  0.60140772-0.43496986 i   &  0.55566949-0.33158637 i &  0.49376269-0.21360446 i \\ \hline
 $\ell=10$ & 9.40544221-0.41934024 i  & 8.15238600-0.31824940 i & 6.68219785-0.17455013 i  &  4.89133776-0.00536578 i  \\ \hline
$\ell=20$ &  18.34294833-0.41859975 i  &  15.88919672-0.31679069 i  & 12.99054975-0.16506712 i  & 9.37995278-0.00005840 i  \\ \hline
$\ell=30$ &  27.28486839-0.41845231 i  & 23.63192200-0.31648560 i  & 19.30804041-0.16194591 i & 13.85644956-0.00000040 i \\ \hline
$\ell=40$ & 36.22794705-0.41839947 i  & 31.37623574-0.31637475 i &  25.62859493-0.16052566 i & 18.33052377-0.0000000025 i  \\ \hline
\end{tabular}
\label{Table}
\end{table}

In the following, we shall consider only the WKB method to study the anomalous behaviour and the critical scalar field mass, which is defined for large-$\ell$ values, where the WKB approximation in general yields accurate values of the QNFs.


\subsection{Critical scalar fields mass}
\label{WKBJ}

In this section, we consider the eikonal limit $\ell \rightarrow \infty$   to estimate the critical scalar field mass for two wormhole geometries, by
considering $\omega_I^\ell=\omega_I^{\ell+1}$ as a proxy for where the transition or critical behaviour occurs \cite{Lagos:2020oek}.


\subsubsection{Generalized Bronnikov-Ellis wormhole}

The maximum of the potential is located at $r=r_0$, which is found from the condition $b(r_0) = r_0$, which yields $r_0= b_0$ or $r_0^{\ast}=0$, and its value is given by
\begin{equation} \label{coa}
V(r^*_0) = -\frac{2M-b_0}{b_0^3} L^2 -\frac{(2M-b_0)(1+m^2 b_0^2)}{b_0^3}  \,,
\end{equation}
while the second derivative of the potential evaluated at $r^*_0$ yields
\begin{equation}
V^{(2)}(r^*_0) = -\frac{2(2M-b_0)(3M-b_0)}{b_0^6}L^2 - \frac{2(2M-b_0)(-2b_0+M(6+m^2 b_0^2))}{b_0^6}   \,.
\end{equation}
For the higher derivatives, we obtain the following expressions
\begin{eqnarray}
\nonumber V^{(3)}(r^*_0)  &=& 0  \,, \\
\nonumber V^{(4)}(r^*_0) &=& -\frac{2 (2M-b_0)(102 M^2 -73 M b_0 +12 b_0^2)}{b_0^9} L^2 -\frac{2(2M -b_0)(36 b_0^2-M b_0 (215+9m^2 b_0^2)+2M^2 (147+11 m^2 b_0^2))}{b_0^9}  \,, \\
\nonumber V^{(5)}(r^*_0) &=& 0  \,, \\
\nonumber  V^{(6)}(r^*_0) &=&  -\frac{2(2M-b_0)(8472 M^3-9406 M^2 b_0+3309 M b_0^2-360b_0^3)}{b_0^{12}} L^2 - \\
&& \frac{2(2M-b_0)(-1440 b_0^3+3 M b_0^2 (4296+75 m^2 b_0^2)+ 8 M^3 (3918+173m^2b_0^2)-2 M^2 b_0 (17836+567 m^2b_0^2))}{b_0^{12}}\,. \label{cof}
\end{eqnarray}
Now, considering $L$ large, the QNFs are approximately given by
\begin{equation}
\omega = \omega_1 L + \omega_0 +\omega_{-1}L^{-1} + \omega_{-2}L^{-2} + \mathcal{O}(L^{-3})~,
\end{equation}
and using (\ref{frecuencia}) we obtain
\begin{eqnarray}
\nonumber  \omega_{1} b_0 &=&  \sqrt{1-2 \frac{M}{b_0}}\,, \\
\nonumber  \omega_{0} b_0 &=&-\frac{i (2 n+1) \sqrt{1-3 \frac{M}{b_0}}}{2}\,, \\
\nonumber  \omega_{-1} b_0&=&  \frac{\left( \frac{M}{b_0} \right)^2 \left(96 b_0^2 m^2-60 n (n+1)+30\right)+ \frac{M}{b_0} \left(-80 b_0^2 m^2+50 n (n+1)-31\right)-8 \left(-2 b_0^2 m^2+n^2+n-1\right)}{32 \left(1-3 \frac{M}{b_0}\right) \sqrt{1-2 \frac{M}{b_0}}} \,,  \\
\nonumber \omega_{-2} b_0&=& - i (2 n+1) \Bigg(12 \left( \frac{M}{b_0} \right)^4 \left(-384 m^2 b_0^2+115 n (n+1)+171\right)+4 \left( \frac{M}{b_0} \right)^3 (1920 m^2 b_0^2-499 n (n+1) \\
\nonumber &&  -723)+ \left( \frac{M}{b_0} \right)^2 \left(-4736 m^2 b_0^2+961 n (n+1)+1585\right)+4 \left(\frac{M}{b_0} \right) \left(5 \left(64 m^2 b_0^2-21\right)-48 n (n+1)\right)  \\
&&  +16 (-8 m^2 b_0^2+ n^2+ n+3)\Bigg)/ \left(512 (1-3 \frac{M}{b_0})^{5/2} (1-2 \frac{M}{b_0}) \right)\,.
\end{eqnarray}

Interestingly, $\omega_{-1}$ and $\omega_{-2}$ diverge when $M/b_0=1/3$ and $1/2$, also the second derivative of the potential $V^{(2)}(r_0^{\ast})$ evaluated at the maximum $r_0^{\ast}$ is zero for those values of $M/b_0$ when $m=0$. Furthermore, for $M/b_0 < 1/3$   $\omega$ is complex, but for $M/b_0 >1/3$, $\omega_0$ and $\omega_{-2}$ become real, therefore $\omega$ is real in the interval $1/3 < M/b_0 < 1/2$ and there is not a critical mass.  Comparing this behaviour with  the more accurate values obtained with the pseudospectral Chebyshev method in Table \ref{Table}, we can observe that for $M/b_0=0.4$ and high values of $\ell$ the QNFs have a very small, but non-null imaginary part which is much smaller than for the other values of $M/b_0=0.1, 0.2, 0.3$. See also Fig.~\ref{omega}.

As we mentioned, in order to estimate the critical scalar field mass we consider $\omega_I^\ell=\omega_I^{\ell+1}$ as a proxy for where the transition or critical behaviour occurs. So, the critical scalar field mass is given by
\begin{equation}
\resizebox{0.9\hsize}{!}{$
  m_c b_0 = \frac{\sqrt{-2  \left( \frac{M}{b_0} \right)^3 (499 n (n+1)+723)+\frac{1}{2}  \left( \frac{M}{b_0} \right)^2 (961 n (n+1)+1585)-6 \left( \frac{ M}{b_0} \right) (16 n (n+1)+35)+8 \left(n^2+n+3\right)+6 \left( \frac{M}{b_0} \right)^4 (115 n (n+1)+171)}}{8 \left(1-3  \frac{M}{b_0} \right) \left( 1-2\frac{M}{b_0}\right)}$} \,,
\end{equation}
valid for $M/b_0 <  1/3$.

\begin{figure}[h!]
\begin{center}
\includegraphics[width=0.5\textwidth]{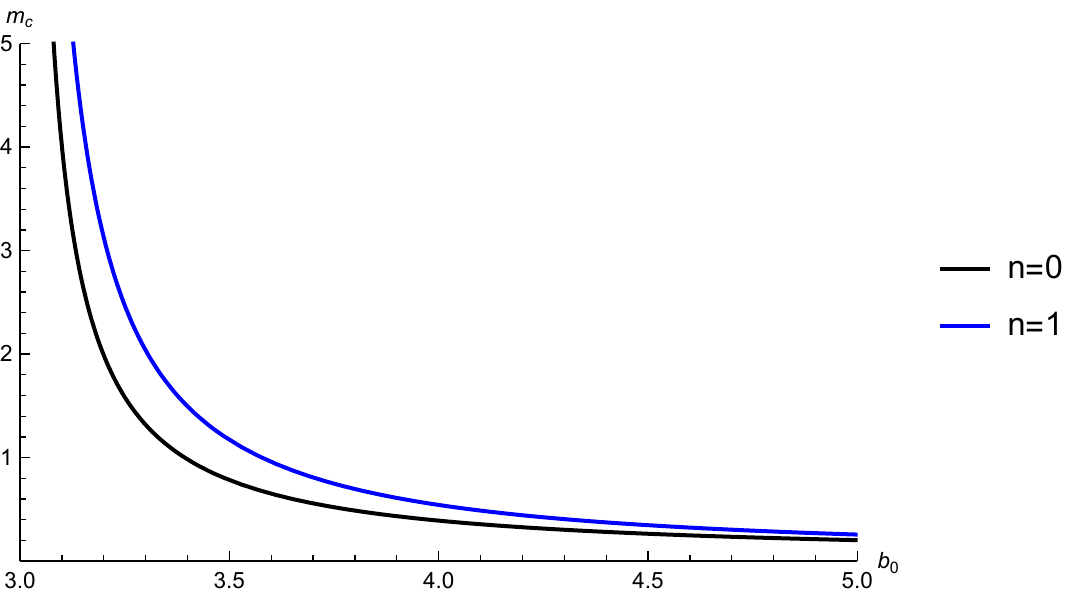}\\
\end{center}
\caption{The behaviour of $m_c$ as a function of $b_0$ for the Generalized Bronnikov-Ellis wormhole.}
\label{masa}
\end{figure}

A general expression for the critical scalar field mass for a generic static and spherically symmetric wormhole (\ref{MorrisThorne-metric}) with arbitrary metric functions $\Phi(r)$ and $b(r)$ is given in Appendix \ref{WKB}. Note that the above expression for $m_c b_0$ diverges when $M/b_0=1/3$, and $1/2$. In Fig. \ref{masa} we plot the behaviour of the critical scalar field mass as a function of $b_0$ for $M=1$ and for the overtone numbers 0, and 1. We observe the critical mass diverges for $b_0=3$. Also, we observe that the critical mass decreases when $b_0$ increases and increases when the overtone number increases.


%
%

\begin{figure}[h!]
\begin{center}
\includegraphics[width=0.4\textwidth]{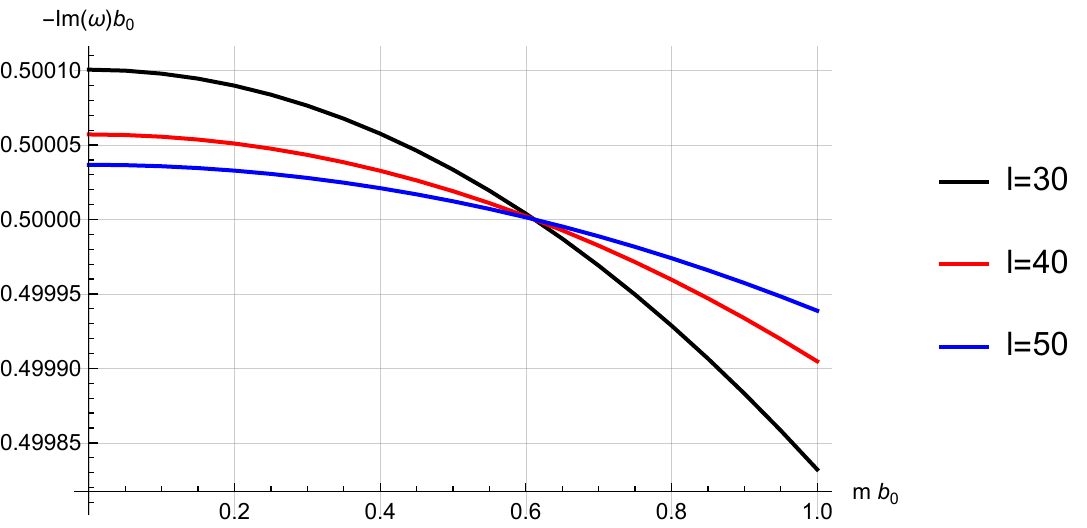}
\includegraphics[width=0.4\textwidth]{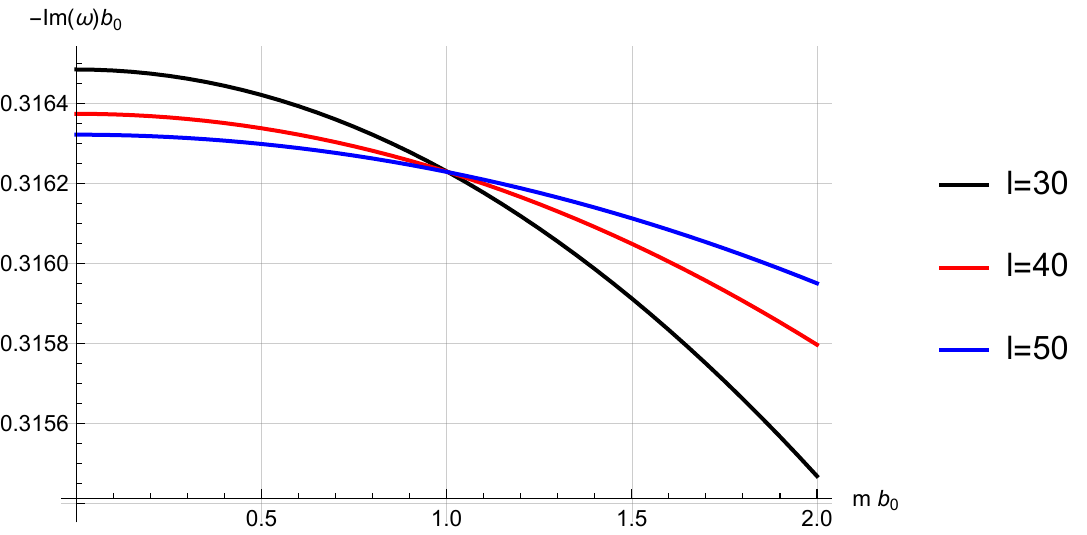}\\
\end{center}
\caption{QNFs as a function of the scalar field mass for the Bronnikov-Ellis wormhole with $\ell = 30, 40, 50$. $M/b_0=0$ left panel, and $M/b_0=0.2$ right panel.}
\label{cmBEW}
\end{figure}

In Fig. \ref{cmBEW} we plot the behaviour of $-Im(\omega)b_0$ as a function of $mb_0$. We can observe an inverted behavior of $\omega_I(\ell)$. For $m>m_c$, $\omega_I$ increases with $\ell$; whereas, for $m<m_c$, $\omega_I$ decreases when $\ell$ increases. The numerical values are in appendix (\ref{NV}), Table \ref{NVBE0} and \ref{NVBE1}. Also, it is possible to observe in such tables that the frequency of the oscillation increases when the angular number increases and when the parameter $mb_0$ increases.

\subsubsection{Morris-Thorne wormhole}

\begin{figure}[h!]
\includegraphics[width=0.4\textwidth]{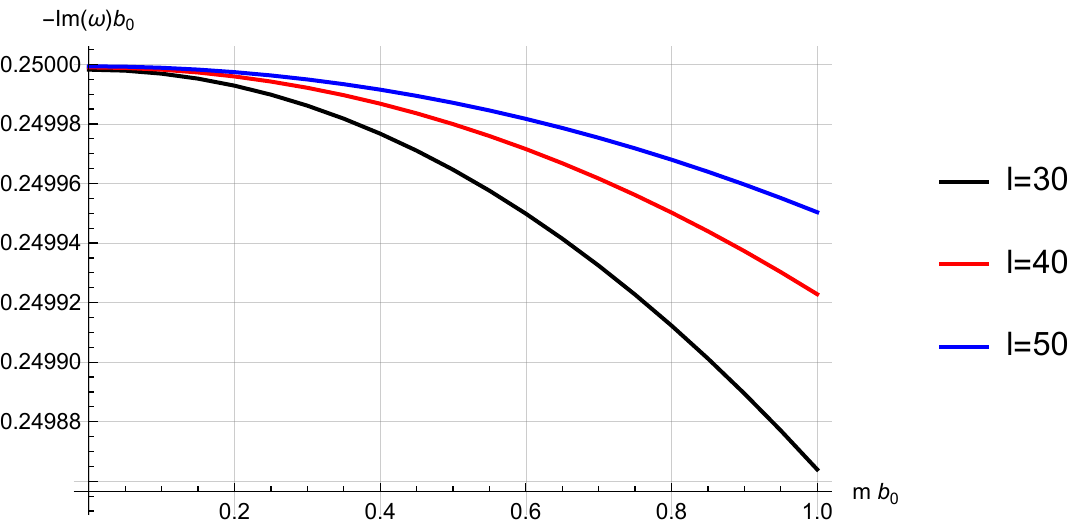}
\includegraphics[width=0.4\textwidth]{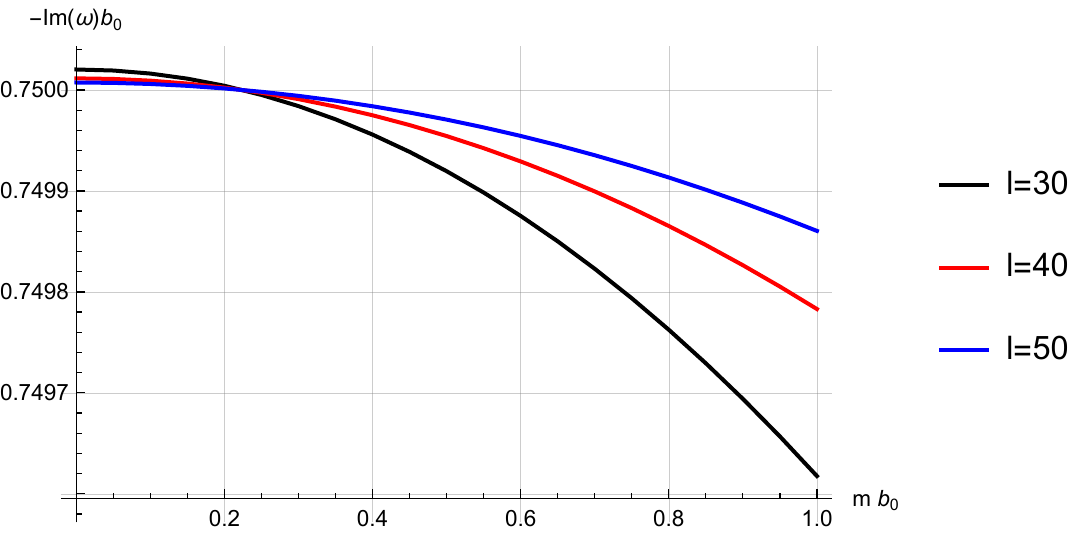}\\
\caption{QNFs as a function of the scalar field mass for the Morris-Thorne wormhole with $\ell = 30, 40, 50$, left panel for $n=0$ and right panel for $n=1$.}
\label{mcMT}
\end{figure}

The maximum of the potential is located at $r= b_0$ or $r_0^{\ast}=0$, and its value is given by
\begin{equation} \label{coa}
V(r^*_0) = \frac{1}{b_0^2}L^2 +m^2+\frac{1}{4 b_0^2}  \,,
\end{equation}
while the second derivative of the potential evaluated at $r^*_0$ yields
\begin{equation}
V^{(2)}(r^*_0) = -\frac{1}{2 b_0^4} L^2 -\frac{5}{32 b_0^4}  \,.
\end{equation}
For the higher derivatives, we obtain the following expressions
\begin{eqnarray}
V^{(3)}(r^*_0)  &=& 0 \,, \\
V^{(4)}(r^*_0) &=&  \frac{21}{16 b_0^6} L^2 +\frac{15}{32 b_0^6}  \,, \\
V^{(5)}(r^*_0) &=& 0 \,, \\
V^{(6)}(r^*_0) &=& - \frac{261}{32 b_0^8} L^2 -\frac{3285}{1024 b_0^8}  \,.
\end{eqnarray}
Then, considering Eq. (\ref{frecuencia}) and the expansion (\ref{aproximado}) for $L$ large the QNFs are approximately given by
\begin{equation}
\omega b_0  =  L -\frac{i (2 n+1)}{4} +\frac{128 b_0^2 m^2-10 n (n+1)+19}{256}  L^{-1} -\frac{i (2 n+1) \left(-2048 b_0^2 m^2+65 n (n+1)-27\right)}{16384}L^{-2} + \mathcal{O}(L^{-3}) \,,
\end{equation}
and the critical scalar field mass is given by
\begin{equation}
    m_c b_0 = \frac{\sqrt{65 n (n+1)-27}}{32 \sqrt{2}}\,.
\end{equation}
Notice that there is a critical scalar field mass only for $n \geq 1$ which increases when the overtone number increases, and the QNFs do not have an anomalous decay rate behaviour for $n=0$. In Fig. \ref{mcMT} we plot the behaviour of $-Im(\omega)b_0$ as a function of $mb_0$. We can observe an inverted behavior of $\omega_I(\ell)$ for $n=1$ (right panel), where for $m>m_c$, $\omega_I$ increases with $\ell$; whereas, for $m<m_c$, $\omega_I$ decreases when $\ell$ increases. However, for the fundamental mode (left panel) there is not an anomalous decay rate behaviour of the QNMs. The numerical values are in appendix (\ref{NV}), Table \ref{NVMT0} and \ref{NVMT1}. Also, it is possible to observe in such tables that the frequency of the oscillation increases when the angular number increases and when the parameter $mb_0$ increases.

\section{Conclusions}

\label{sec5}

In this work we considered the Generalized Bronnikov-Ellis wormhole and the   Morris-Thorne wormhole and we studied the propagation of massive scalar fields. We showed for the Generalized Bronnikov-Ellis wormhole that there is an  anomalous decay rate of the quasinormal modes depending on a critical value of the mass of the scalar field which increases as the overtone number is increasing. In particular, we found that  for $b_0< 3M$, where $b_0$  is the scalar charge forming the wormhole, that there is no a critical mass for the scalar field and the behaviour of the quasinormal frequencies  is always anomalous. On the other hand, for $b_0 > 3M$ there is a critical mass $m_c$ of the scalar field and the quasinormal frequencies  present an anomalous behaviour for $m<m_c$ which changes for $m>m_c$. Also, the critical mass diverges for $b_0 = 3M$. Also, we showed that the propagation of massive scalar fields is stable in this wormhole background with a frequency of the oscillation increasing  when the angular number increases and when the parameter $mb_0$ increases.

On the other hand,  for the  Morris-Thorne wormhole the anomalous decay rate of the quasinormal modes is not present for the fundamental mode $n$. However, for $n \geq 1$, there is a range of values of the scalar field mass where the anomalous decay rate for the quasinormal modes is present, depending on a critical scalar field mass. Also the propagation of massive scalar fields is stable with a frequency of the oscillation increasing when the angular number increases and when the parameter $mb_0$ increases.

Furthermore, we employed the pseudospectral method that allows us to calculate the values of the QNFs for scalar field for large as well for small values for the angular number $\ell$, and the semi-analytical WKB method that only works well for high values of $\ell$. In particular, it should be mentioned that the WKB method at sixth-order employed in \cite{Konoplya:2019hlu} certainly allowed to diminish the relative error of the QNFs for several cases by quite a few times or even orders. However, the WKB approach did not allow one to compute $n >> \ell$ modes with satisfactory accuracy in many circumstances as  in \cite{Konoplya:2019hlu}. Thus, the inclusion of the computation of the QNMs using the pseudospectral method is novel by itself.

It would be interesting to extent this work to the study of wormholes in AdS spacetimes.   Wormholes in AdS spacetimes where discussed in \cite{Maldacena:2004rf}, in an attempt to yield some information about the physics of closed Universes. Such discussion is connected with the physics of inflation, and its connection with vacuum decay. A unique realization of such ideas is baby-Universe formation by quantum tunneling which eventually disconnect from the parent spacetime \cite{Giddings:1988wv}. Recently, these ideas of connecting the physics of wormhole spacetimes to baby-Universes were revisited in \cite{Marolf:2020xie}, using features associated with a negative cosmological constant and asymptotically AdS boundaries. The presence of the two AdS boundaries will make the scattered waves to be trapped in the wormhole photosphere and it would be interesting to study there effect on the quasinormal modes and quasinormal frequencies.


\section*{Acknowlegements}
\noindent
This work is partially supported by ANID Chile through FONDECYT Grant Nº 1220871  (P.A.G., and Y. V.).
\\
Á. R. is funded by the María Zambrano contract ZAMBRANO21-25 (Spain).

\newpage

\appendix

\section{WKB method}
\label{WKB}

In this appendix we show the expressions of the QNFs and the critical scalar field mass obtained using the third order WKB approximation for a generic static and spherically symmetric wormhole (\ref{MorrisThorne-metric}) with arbitrary metric functions $\Phi(r)$ and $b(r)$ for which the WKB approximation is applicable. The maximum of the potential (\ref{pot}) is located at $r=r_0$, which is found from the condition $b(r_0) = r_0$.
\newline

For large $L$, the QNFs are approximately given by
\begin{equation}
\omega = \omega_1 L + \omega_0 +\omega_{-1}L^{-1} + \omega_{-2}L^{-2} + \mathcal{O}(L^{-3}) \,,
\end{equation}
where, for $n=0$
\begin{eqnarray}
\omega_1 &=& e^{\Phi(r_0)}/r_0\,,     \\
\omega_0 &=& -\frac{i e^{\Phi(r_0)} \sqrt{(-1+b'(r_0))  (-1+r_0 \Phi'(r_0))}}{ 2 \sqrt{2} r_0}\,,  \\
\nonumber \omega_{-1} &=&  e^{\Phi(r_0)} \Big(  -9-32r_0^2 m^2 +9 b'(r_0) +r_0 b''(r_0) + r_0 (6+32r_0^2m^2 -6b'(r_0) -r_0 b''(r_0)) \Phi'(r_0)  \\
&& -6r_0^2 (-1+b'(r_0)) \Phi'(r_0)^2 -3r_0^2 (-1+b'(r_0))\Phi''(r_0) \Big)/\left(  64 r_0 (-1+r_0 \Phi'(r_0)) \right)\,, \\
\nonumber \omega_{-2} &=& -i e^{\Phi(r_0)} \Big( 20 r_0^3 \Phi ^{(3)}\left(r_0\right) \left(b'\left(r_0\right)-1\right){}^2+21 r_0^4 \left(b'\left(r_0\right)-1\right){}^2 \Phi ''\left(r_0\right){}^2-28 r_0^4 \left(b'\left(r_0\right)-1\right){}^2 \Phi '\left(r_0\right){}^4+13 b'\left(r_0\right){}^2 \\
\nonumber &&  +r_0 \left(4 r_0 b^{(3)}\left(r_0\right)+b''\left(r_0\right) \left(r_0 b''\left(r_0\right)+46\right)\right)+6 r_0^2 \left(b'\left(r_0\right)-1\right) \Phi ''\left(r_0\right) \left(r_0 b''\left(r_0\right)-b'\left(r_0\right)+1\right)   \\
\nonumber && -4 r_0^3 \left(b'\left(r_0\right)-1\right) \Phi '\left(r_0\right){}^3 \left(5 r_0 b''\left(r_0\right)+14 b'\left(r_0\right)-14\right)+\Phi '\left(r_0\right){}^2 \Big(r_0^2 \Big(216 b'\left(r_0\right){}^2+r_0 \Big(4 r_0 b^{(3)}\left(r_0\right)  \\
\nonumber && +b''\left(r_0\right) \left(r_0 b''\left(r_0\right)+12\right)\Big)+4 b'\left(r_0\right) \left(-r_0 \left(r_0 b^{(3)}\left(r_0\right)+3 b''\left(r_0\right)\right)+64 m^2 r_0^2-108\right)-256 m^2 r_0^2+216\Big)  \\
\nonumber &&  -60 r_0^4 \left(b'\left(r_0\right)-1\right){}^2 \Phi ''\left(r_0\right)\Big)+\Phi '\left(r_0\right) \Big(-\Phi ^{(3)}\left(r_0\right) \left(20 r_0^4 \left(b'\left(r_0\right)-1\right){}^2\right)-\Phi ''\left(r_0\right) \Big(6 r_0^3 \left(b'\left(r_0\right)-1\right) (r_0 b''\left(r_0\right)  \\
\nonumber  &&   -18 b'\left(r_0\right)+18)\Big)-2 r_0 \Big(62 b'\left(r_0\right){}^2+r_0 \left(4 r_0 b^{(3)}\left(r_0\right)+b''\left(r_0\right) \left(r_0 b''\left(r_0\right)+39\right)\right)+b'\left(r_0\right) \Big(-4 r_0^2 b^{(3)}\left(r_0\right)  \\
\nonumber   &&   -39 r_0 b''\left(r_0\right)+256 m^2 r_0^2-124\Big)-256 m^2 r_0^2+62\Big)\Big)+b'\left(r_0\right) \left(-4 r_0^2 b^{(3)}\left(r_0\right)-46 r_0 b''\left(r_0\right)+256 m^2 r_0^2-26\right)   \\
&&  -256 m^2 r_0^2+13   \Big)/\Big( 1024 \sqrt{2} r_0 (-1+ r_0 \Phi'(r_0))^2 \sqrt{(-1+b'(r_0)) (-1+r_0 \Phi'(r_0))}  \Big)\,.
\end{eqnarray}
The critical scalar field mass $m_c$ is given by
\begin{equation}\label{mc}
m_c = \sqrt{P} / Q \,,
\end{equation}
where
\begin{eqnarray}
\nonumber P &=&  -60 r_0^3 \Phi ^{(3)}\left(r_0\right) \left(b'\left(r_0\right)-1\right){}^2-63 r_0^4 \left(b'\left(r_0\right)-1\right){}^2 \Phi ''\left(r_0\right){}^2+84 r_0^4 \left(b'\left(r_0\right)-1\right){}^2 \Phi '\left(r_0\right){}^4-39 b'\left(r_0\right){}^2 \\
\nonumber && -3 r_0 \left(4 r_0 b^{(3)}\left(r_0\right)+b''\left(r_0\right) \left(r_0 b''\left(r_0\right)+46\right)\right)-18 r_0^2 \left(b'\left(r_0\right)-1\right) \Phi ''\left(r_0\right) \left(r_0 b''\left(r_0\right)-b'\left(r_0\right)+1\right)  \\
\nonumber && +12 r_0^3 \left(b'\left(r_0\right)-1\right) \Phi '\left(r_0\right){}^3 \left(5 r_0 b''\left(r_0\right)+14 b'\left(r_0\right)-14\right)+\Phi '\left(r_0\right){}^2 \Big(180 r_0^4 \left(b'\left(r_0\right)-1\right){}^2 \Phi ''\left(r_0\right)+ \\
\nonumber &&  3 r_0^2 \Big(-216 b'\left(r_0\right){}^2-r_0 \left(4 r_0 b^{(3)}\left(r_0\right)+b''\left(r_0\right) \left(r_0 b''\left(r_0\right)+12\right)\right)+4 b'\left(r_0\right) \left(r_0^2 b^{(3)}\left(r_0\right)+3 r_0 b''\left(r_0\right)+108\right)  \\
\nonumber &&  -216\Big)\Big)+\Phi '\left(r_0\right) \Big(60 r_0^4 \Phi ^{(3)}\left(r_0\right) \left(b'\left(r_0\right)-1\right){}^2+18 r_0^3 \left(b'\left(r_0\right)-1\right) \Phi ''\left(r_0\right) \left(r_0 b''\left(r_0\right)-18 b'\left(r_0\right)+18\right)+ \\
\nonumber  && 6 r_0 \left(62 b'\left(r_0\right){}^2+r_0 \left(4 r_0 b^{(3)}\left(r_0\right)+b''\left(r_0\right) \left(r_0 b''\left(r_0\right)+39\right)\right)-b'\left(r_0\right) \left(4 r_0^2 b^{(3)}\left(r_0\right)+39 r_0 b''\left(r_0\right)+124\right)+62\right)\Big)  \\
&& +6 b'\left(r_0\right) \left(2 r_0^2 b^{(3)}\left(r_0\right)+23 r_0 b''\left(r_0\right)+13\right)-39  \,, \\
Q   &=& 16 \sqrt{3} r_0 \sqrt{(-1+b'(r_0))(-1+r_0 \Phi'(r_0))^2} \,.
\end{eqnarray}


\newpage

\section{Numerical values}
\label{NV}

In the following tables we show the numerical values of the QNFs obtained with the WKB method used to plot the Figs. \ref{cmBEW} and \ref{mcMT}
\begin{table}[h]
\centering
\caption{Fundamental QNFs $\omega b_0$ for scalar perturbations for the Bronnikov-Ellis wormhole, for several values of the angular momentum and the mass of the scalar field using the six order WKB approximation. }
\label{NVBE0}
\begin{tabular}{|c|c|c|c|}
\hline
$\ell$ &  $m b_0=0$ & $m b_0=0.2$ & $m b_0=0.4$ \\
\hline
  10  & 10.51183943-0.50083950 i & 10.51373757-0.50074908 i & 10.51942994-0.50047811 i  \\
  20  & 20.50608858-0.50022226 i & 20.5070633-0.50019848 i & 20.50998717-0.50012717 i  \\
  30   & 30.50409562-0.50010061 i & 30.50475109-0.50008986 i  & 30.5067174-0.50005763 i  \\
    40  & 40.50308525-0.50005710 i  & 40.50357896-0.50005101 i & 40.50506006-0.50003272 i  \\

   50  & 50.50247464-0.50003674 i & 50.50287062-0.50003282 i & 50.50405854-0.50002106 i  \\
\hline
\hline
$\ell$ &  $m b_0=0.6$ & $m b_0=0.8$ & $m b_0=1$ \\
\hline
  10  & 10.52891046-0.50002747 i & 10.54216899-0.49939860 i &  10.55919142-0.49859352 i \\
  20  & 20.51485937-0.50000840 i & 20.52167851-0.49984225 i & 20.53044265-0.49962887 i  \\
  30   &  30.50999431-0.50000392 i  & 30.51458139-0.49992876 i  &  30.52047806-0.49983217 i \\
    40  & 40.50752843-0.50000225 i  & 40.51098391-0.49995960 i &  40.51542622-0.49990478 i \\
   50  & 50.50603835-0.50000146 i & 50.50880995-0.49997402 i & 50.51237321-0.49993875 i  \\
\hline
\end{tabular}
\end{table}
\begin{table}[h]
\centering
\caption{Fundamental QNFs $\omega b_0$ for scalar perturbations for the Generalized Bronnikov-Ellis wormhole with $M/b_0 =0.2$, for several values of the angular momentum and the mass of the scalar field using the six order WKB approximation. }
\label{NVBE1}
\begin{tabular}{|c|c|c|c|}
\hline
$\ell$ &  $m b_0=0$ & $m b_0=0.2$ & $m b_0=0.4$ \\
\hline
  10  & 8.152424282-0.31819085 i & 8.15389325-0.31810533 i & 8.158298582-0.31784900 i \\
  20  & 15.88919723-0.31678931 i & 15.88995206-0.31676678 i & 15.89221637-0.31669919 i  \\
  30   & 23.63192203-0.31648546 i & 23.6324297-0.31647527 i  & 23.63395265-0.31644471 i  \\
    40  & 31.37623575-0.31637472 i & 31.37661815-0.31636894 i & 31.37776534-0.31635160 i  \\
   50  & 39.12120109-0.31632253 i & 39.1215078-0.31631881 i & 39.12242793-0.31630765 i  \\
\hline
\hline
$\ell$ &  $m b_0=0.6$ & $m b_0=0.8$ & $m b_0=1$ \\
\hline
  10  & 8.165635572-0.31742255 i & 8.175896403-0.31682714 i & 8.18907019-0.31606434 i  \\
  20  & 15.8959895-0.31658661 i & 15.90127037-0.31642910 i  & 15.9080575-0.31622679 i  \\
  30   &  23.63649067-0.31639378 i  & 23.64004345-0.31632250 i  & 23.64461052-0.31623089 i  \\
    40  &  31.37967722-0.31632270 i & 31.38235367-0.31628225 i & 31.38579448-0.31623025 i  \\
   50  & 39.12396142-0.31628906 i & 39.12610821-0.31626304 i & 39.1288682-0.31622958 i  \\
\hline
\end{tabular}
\end{table}
\begin{table}[h]
\centering
\caption{Fundamental QNFs $\omega b_0$ for scalar perturbations for the Morris-Thorne wormhole, for several values of the angular momentum and the mass of the scalar field using the six order WKB approximation. }
\label{NVMT0}
\begin{tabular}{|c|c|c|c|}
\hline
$\ell$ &  $m b_0=0$ & $m b_0=0.2$ & $m b_0=0.4$ \\
\hline
  10  & 10.49516376-0.24998509 i & 10.49706815-0.24993973 i & 10.50277924-0.24980382 i \\
  20  & 20.49752288-0.24999608 i & 20.49849844-0.24998418 i &  20.50142483-0.24994850 i \\
  30   & 30.49833504-0.24999823 i & 30.49899077-0.24999285 i  & 30.50095785-0.24997673 i  \\
    40  & 40.49874614-0.24999900 i & 40.49923996-0.24999595 i &  40.50072139-0.24998680 i \\
   50  & 50.49899443-0.24999935 i & 50.49939047-0.24999739 i & 50.50057856-0.24999151 i  \\
\hline
\hline
$\ell$ &  $m b_0=0.6$ & $m b_0=0.8$ & $m b_0=1$ \\
\hline
  10  & 10.51229085-0.24957780 i & 10.5255927-0.24926239 i & 10.54267046-0.24885862 i  \\
  20  & 20.50630124-0.24988906 i & 20.51312626-0.24980592 i & 20.52189795-0.24969915 i  \\
  30   &  30.50423604-0.24994987 i  & 30.50882491-0.24991227 i  & 30.51472388-0.24986396 i  \\
    40  & 40.50319031-0.24997156 i  & 40.50664655-0.24995024 i &  40.51108985-0.24992282 i \\
   50  & 50.50255865-0.24998171 i & 50.50533064-0.24996799 i & 50.50889441-0.24995035 i  \\
\hline
\end{tabular}
\end{table}
\begin{table}[h]
\centering
\caption{First overtone $n=1$ QNFs $\omega b_0$ for scalar perturbations for the Morris-Thorne wormhole, for several values of the angular momentum and the mass of the scalar field using the six order WKB approximation. }
\label{NVMT1}
\begin{tabular}{|c|c|c|c|}
\hline
$\ell$ &  $m b_0=0$ & $m b_0=0.2$ & $m b_0=0.4$ \\
\hline
  10  & 10.48772523-0.75017126 i & 10.48962235-0.75003559 i & 10.49531168-0.74962901 i \\
  20  & 20.49371216-0.75004489 i & 20.49468674-0.75000922 i &  20.49761021-0.74990225 i \\
  30   & 30.49577365-0.75002028 i & 30.49642907-0.75000416 i  & 30.49839526-0.74995581 i  \\
    40  & 40.49681716-0.75001150 i & 40.49731086-0.75000236 i &  40.4987919-0.74997492 i \\
   50  & 50.49744742-0.75000740 i & 50.49784339-0.75000151 i & 50.49903128-0.74998387 i  \\
\hline
\hline
$\ell$ &  $m b_0=0.6$ & $m b_0=0.8$ & $m b_0=1$ \\
\hline
  10  & 10.50478719-0.74895283 i & 10.51803886-0.74800922 i & 10.53505271-0.74680121 i  \\
  20  & 20.50248173-0.74972407 i & 20.50929992-0.74947483 i & 20.51806286-0.74915474 i  \\
  30   &  30.50167197-0.74987524 i  & 30.50625876-0.74976249 i  & 30.51215506-0.74961760 i  \\
    40  & 40.50126019-0.74992922 i  & 40.50471554-0.74986525 i &  40.5091577-0.74978302 i \\
   50  & 50.50101104-0.74995447 i & 50.50378258-0.74991331 i & 50.50734576-0.74986041 i  \\
\hline
\end{tabular}
\end{table}

\end{document}